\RequirePackage{fix-cm}
\documentclass[smallextended]{svjour3}       
\smartqed  
\usepackage{graphicx}
\usepackage{multirow}
\usepackage{epsfig}
\journalname{}

\title{Parity proofs of the Kochen-Specker theorem based on the 24 rays of Peres}

\author{Mordecai Waegell and P.K. Aravind }

\authorrunning{M.Waegell, P.K. Aravind}

\institute{M.Waegell, P.K. Aravind \at
Physics Department, Worcester Polytechnic Institute, Worcester, MA 01609, U.S.A.\\
\email{caiw@wpi.edu, paravind@wpi.edu}
}

\date{\today}

\begin{document}
\maketitle
\begin{abstract}
 A diagrammatic representation is given of the 24 rays of Peres that makes it easy to pick out all the 512 parity proofs of the Kochen-Specker theorem contained in them. The origin of this representation in the four-dimensional geometry of the rays is pointed out.\\
\end{abstract}

\begin{figure}[htp]
\begin{center}
\includegraphics[width=0.80\textwidth]{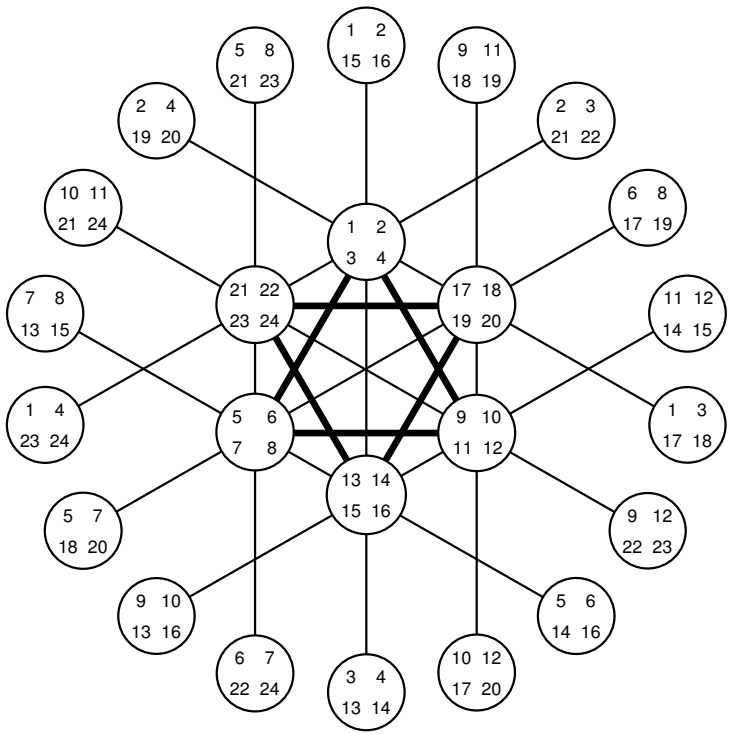}
\end{center}
\caption{The 24 bases formed by the 24 Peres rays}
\label{hex1}
\end{figure}

Some time back Peres \cite{Peres1991} gave a proof of the Kochen-Specker (K-S) theorem \cite{KS1967} using 24 rays in a real four-dimensional Hilbert space. His proof was much simpler than the original proof of Kochen and Specker, which used 117 rays (or unoriented directions) in ordinary three-dimensional space. Peres's proof was simplified by Kernaghan \cite{Kernaghan1994} and Cabello et al \cite{Cabello1996}, who showed that the Peres rays contain several subsets of 20 and 18 rays that provide transparent ``parity'' proofs of the theorem. The existence of further parity proofs involving 22 and 24 rays was pointed out by Pavi{\v c}i{\'c} et al \cite{Pavicic2010}. This paper presents a diagram (Fig.\ref{hex1}) that permits a simple visualization of the $2^{9}=512$ parity proofs in this system. We invite the reader to solve a Sudoku-like puzzle, with easily stated rules, whose solutions yield the parity proofs. This puzzle can be attempted even by non-physicists. We give a solution to the puzzle in the form of a simple set of rules for generating all the parity proofs. We then discuss the four-dimensional geometry of the Peres rays that underlies Fig.\ref{hex1} and helps explain the rules for generating the parity proofs. Although no new results are presented in this paper, we believe it still serves a useful purpose because it gives a unified account of all the parity proofs in this system obtained by many authors over a period of many years. The interest of this demonstration in relation to ongoing work on the Kochen-Specker theorem is discussed in the concluding section of this paper.\\ \\

{\bf The Peres rays and their bases}\\

The 24 Peres rays arise as the simultaneous eigenstates of the six triads of commuting observables for a pair of qubits shown in Table \ref{Rays}. The rays form 24 bases of four mutually orthogonal rays that are shown within the circular boxes in Fig.\ref{hex1}. Each ray occurs in four bases and its nine companions in these bases (three of which occur twice with it and six just once) are the only other rays it is orthogonal to. The six inner bases divide into two groups of three, each at the corners of an equilateral triangle. One group contains the rays 1-12 and the other the rays 13-24. Each outer basis lies on a line passing through two inner bases and one other outer basis. The two outer bases along a line consist of complementary mixes of the rays in the two inner bases between them. For example, the outer bases 1 2 15 16 and 3 4 13 14 are ``hybrids'' of the inner bases 1 2 3 4 and 13 14 15 16. \\

\begin{table}[ht]
\begin{center} 
\begin{tabular}{|c | c c c c |} 
\hline 
Observables & \multicolumn{4}{|c|}{Rays}  \\
\hline
$\sigma_z^1$, $\sigma_z^2$, $\sigma_z^1$$\sigma_z^2$            & $1=2000$ & $2=0200$ & $3=0020$ & $4=0002$ \\
$\sigma_x^1$, $\sigma_x^2$, $\sigma_x^1$$\sigma_x^2$            & $5=1111$ & $6=11\overline{1}\overline{1}$ & $7=1\overline{1}1\overline{1}$ & $8=1\overline{1}\overline{1}1$ \\
$\sigma_z^1$$\sigma_x^2$, $\sigma_x^1$$\sigma_z^2$, $\sigma_y^1$$\sigma_y^2$   & $9=1\overline{1}\overline{1}\overline{1}$ & $10=1\overline{1}11$& $11=11\overline{1}1$  & $12=111\overline{1}$ \\
\hline
$\sigma_z^1$, $\sigma_x^2$, $\sigma_z^1$$\sigma_x^2$            & $13=1100$ & $14=1\overline{1}00$ & $15=0011$ & $16=001\overline{1}$ \\
$\sigma_x^1$, $\sigma_z^2$, $\sigma_x^1$$\sigma_z^2$            & $17=0101$ & $18=010\overline{1}$ & $19=1010$ & $20=10\overline{1}0$ \\
$\sigma_z^1$$\sigma_z^2$, $\sigma_x^1$$\sigma_x^2$, $\sigma_y^1$$\sigma_y^2$  & $21=100\overline{1}$ & $22=1001$ & $23=01\overline{1}0$ & $24=0110$ \\
\hline
\end{tabular}
\end{center}
\caption{The Peres rays. The first column shows six triads of commuting observables for a pair of qubits. Each row of the second column shows the four rays that arise as the simultaneous eigenstates of the triad of observables to its left. The rays are numbered 1 to 24, with commas omitted between their components and a bar over a number indicating its negative.}
\label{Rays}
\end{table}

{\bf The parity proof puzzle}\\

A proof of the K-S theorem with the Peres rays requires showing that it is impossible to assign the value 0 or 1 to each of the rays in such a way that each basis, or circular box in Fig.\ref{hex1}, contains exactly one 1 and three 0's in it (see Appendix 1 for a quick explanation of the K-S theorem and why it can be proved in this way). The impossibility of this assignment can be demonstrated by finding $R$ rays that form $B$ complete bases, with $B$ odd, in such a way that each ray occurs an even number of times over these bases. The reason is that, on the one hand, the number of 1's over all the bases must be odd (because there can be only one 1 in a basis and the total number of bases is odd) while, on the other, the number of 1's over all the bases must be even (because each ray assigned the value 1 is repeated an even number of times). The impossibility of the value assignment for this limited set of rays and bases implies that it is impossible for the entire set as well, and the K-S theorem is proved. We will say that any set of $R$ rays and $B$ bases that gives rise to this even-odd contradiction provides a $R$-$B$ parity proof of the K-S theorem. It is easy to see that a $R$-$B$ parity proof has $2R-2B$ rays of multiplicity two and $2B-R$ rays of multiplicity four (the multiplicity of a ray is the number of times it occurs over the bases). The Peres rays contain four different types of parity proofs, whose characteristics are listed in Table \ref{Summary}. The total number of proofs of all four kinds is 512. The reader is now invited to lay aside this article and see how many of these proofs s/he can pick out from Fig.\ref{hex1}. For example, finding a 18-9 proof involves picking out 9 circles (bases) containing 18 different numbers in such a way that each of the numbers occurs twice among those circles. Happy puzzling!\\

\begin{table}[ht]
\centering 
\begin{tabular}{|c |c |c |c|} 
\hline 
& \multicolumn{3}{|c|}{Number of...}\\
\hline
R-B & Rays of multiplicity 2 & Rays of multiplicity 4 & Proofs\\
\hline
18-9 & 18 & 0  & 16\\
20-11 &  18 &  2   & 240\\
22-13&  18 & 4   &   240\\
24-15&  18 & 6   &  16 \\
\hline
\end{tabular}
\caption{Parity proofs contained within the 24-24 system of Peres rays and their bases.}
\label{Summary} 
\end{table}

{\bf ... and its solution}\\

Note that if one finds a 18-9 proof, the 15 bases left out of it automatically give a 24-15 proof. Similarly, if one finds a 20-11 proof, the 13 bases left out of it give a 22-13 proof. Thus, if one finds all the 18-9 and 20-11 proofs, their ``basis-complements'' in the full 24-24 will yield all the 22-13 and 24-15 proofs. The equality in the number of 18-9 and 24-15 proofs (as well as 20-11 and 22-13 proofs) in Table \ref{Summary} should now be clear. \\

A recipe for generating all the 18-9 proofs and some of the 20-11 proofs is as follows: \\

\begin{figure}[htp]
\begin{center}
\includegraphics[width=0.80\textwidth]{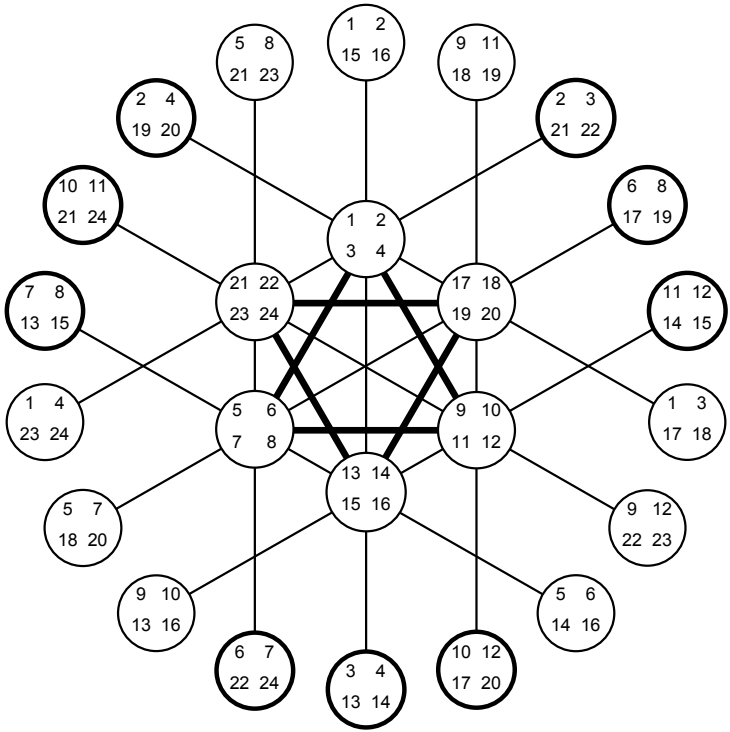}
\end{center}
\caption{A 18-9 parity proof (bold circles) and its complementary 24-15 proof (light circles)}
\label{hex2}
\end{figure}

\begin{figure}[htp]
\begin{center}
\includegraphics[width=0.80\textwidth]{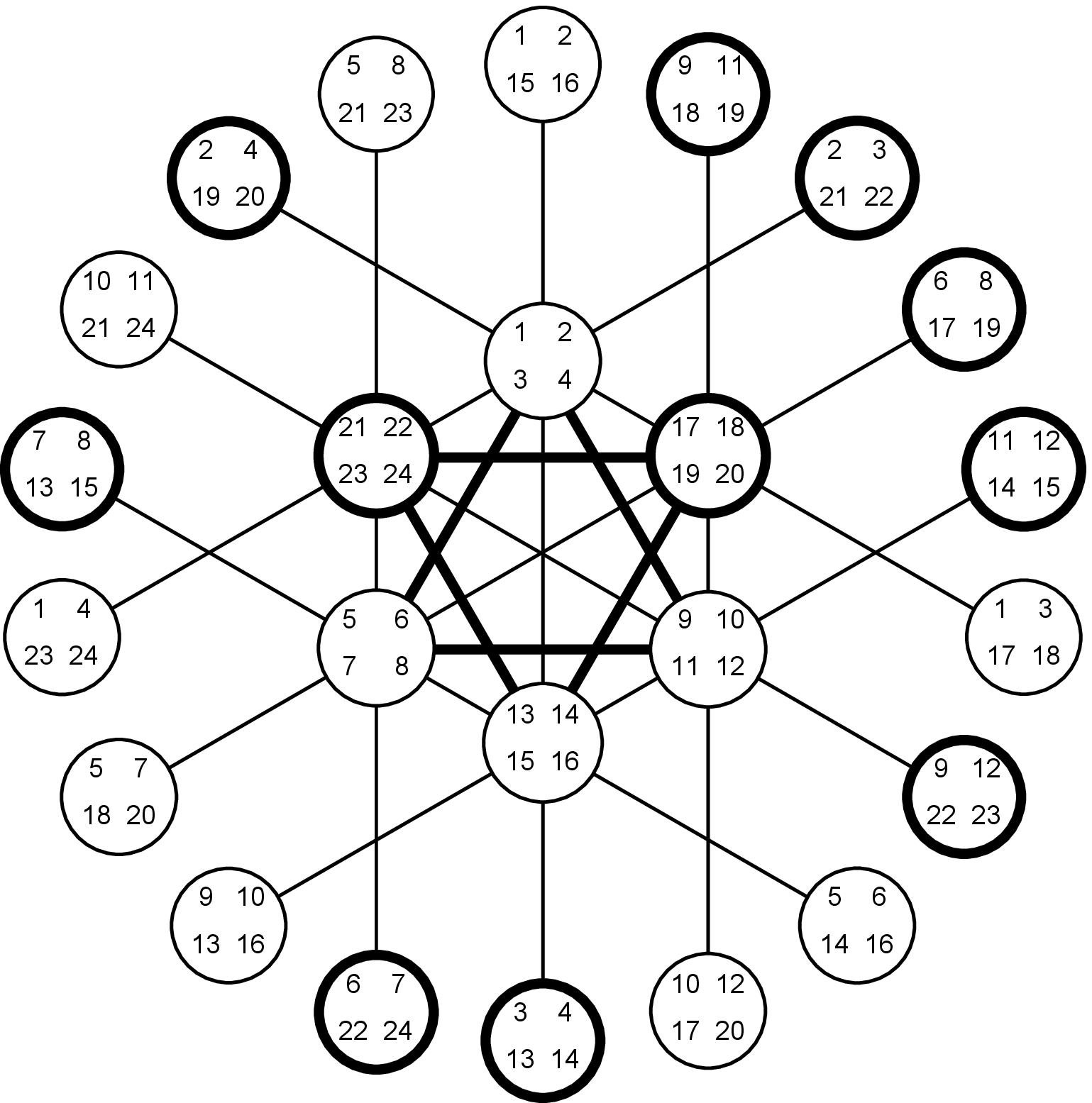}
\end{center}
\caption{A 20-11 parity proof of the first type (bold circles) and its complementary 22-13 proof (light circles)}
\label{hex3}
\end{figure}

(i) Pick one ray from each of the groups 1-4, 5-8 and 9-12 (these three rays can be picked in a total of $4^{3} = 64$ ways). \\
(iia) In 16 of the 64 cases, the 9 outer circles not containing any of the rays picked in step (i) yield a 18-9 proof. For example, picking rays 1,5 and 9 yields the 18-9 proof shown in Fig.\ref{hex2}.\\
(iib) In the remaining 48 out of 64 cases, the 9 outer circles picked in step (iia) do not yield a 18-9 proof. However they can be supplemented by two inner circles to yield a 20-11 proof. The two inner circles that must be picked are the ones that do not involve any of the rays picked in step (i) or their common companion in three of the outer circles. For example, if rays 1,5 and 10 are picked in step (i), their common companion in the outer circles is ray 16, and following the stated procedure leads to the 20-11 proof shown in Fig.\ref{hex3}.\\
(iii) Repeat steps (i) and (ii), but now picking one ray from each of the groups 13-16, 17-20 and 21-24. This yields the same 18-9 proofs as before, but 48 new 20-11 proofs.\\

The above construction yields all the 18-9 proofs and 96 of the 20-11 proofs.  The remaining 20-11 proofs can be obtained as follows:\\

\begin{figure}[htp]
\begin{center}
\includegraphics[width=0.80\textwidth]{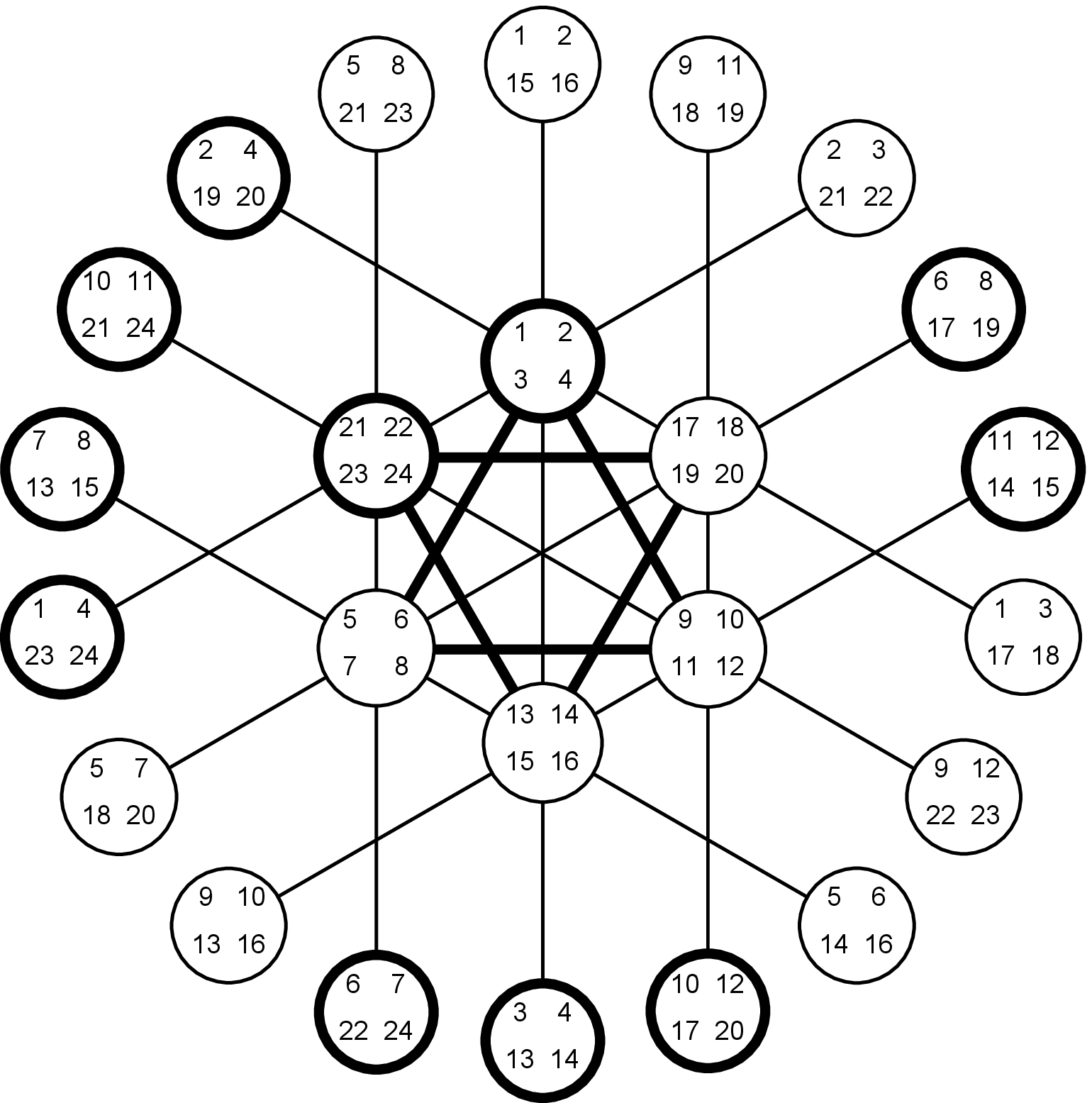}
\end{center}
\caption{A 20-11 parity proof of the second type (bold circles) and its complementary 22-13 proof (light circles)}
\label{hex4}
\end{figure}

(iv) Consider any of the 18-9 proofs obtained earlier. Dropping any of the outer circles from this proof and adding the three other circles (two inner and one outer) on the same line as it gives a 20-11 proof. For example, if one begins from the 18-9 proof of Fig.\ref{hex2}, drops the basis 2 3 21 22 and adds the three bases 1 2 3 4, 21 22 23 24 and 1 4 23 24, one gets the 20-11 proof shown in Fig.\ref{hex4}. The number of such proofs is $16\times9=144$. This brings the total number of 20-11 proofs up to 240.\\

{\bf Geometry of the Peres rays: inner machinery of the puzzle}\\

Let us interpret the Peres rays as vectors in four-dimensional Euclidean space with the components given in Table \ref{Rays}. Then vectors 1-12, together with their negatives, define the vertices of a four-dimensional regular polytope known as a 24-cell \cite{Coxeter} while vectors 13-24, together with their negatives, define the vertices of another 24-cell rotated with respect to the first. If the vectors are normalized to unit length, then the vertices of the two 24-cells lie on a sphere of unit radius centered at the origin. Each 24-cell has 24 vertices and 24 regular octahedra for its boundaries (the latter accounting for its name). The vertices of either 24-cell are in the directions of the cell centers of the other from their common center. Because of this, the two cells are said to be the duals of each other. The Peres rays 1-12 correspond to antipodal vertex pairs of one cell (which we will term Cell A), and rays 13-24 to antipodal vertex pairs of the other cell (which we will term Cell B).

The 12 vertex-pairs of any 24-cell split up into three groups of four, with the vertex pairs in any group being in mutually orthogonal directions from its center and constituting the vertices of a simpler four-dimensional figure known as a cross polytope (or 16-cell). The vertex pairs that make up a cross polytope define a basis of rays. Cell A is made up of the three cross polytopes (or bases) at the corners of one of the equilateral triangles at the center of Fig. \ref{hex1} and Cell B of the cross polytopes at the corners of the other triangle. There is a special linear transformation, known as a Clifford displacement, that takes the vertices of a 24-cell into those of its dual. For our pair of 24-cells it is given by \cite{Coxeter}

\begin{equation}
x_{1}'=\frac{1}{\sqrt{2}}(x_{1}-x_{2}),x_{2}'=\frac{1}{\sqrt{2}}(x_{1}+x_{2}),x_{3}'=\frac{1}{\sqrt{2}}(x_{3}-x_{4}),x_{4}'=\frac{1}{\sqrt{2}}(x_{3}+x_{4})
\label{Clifford}
\end{equation}
Applying this transformation to the normalized ray $i$ of Cell A takes it into the normalized ray $i+12$ of Cell B.\\

It is an interesting geometrical fact that a pair of dual 24-cells whose vertices lie on a common sphere give rise to 18 cross polytopes whose vertices arise equally from the two of them. For our 24-cells, these "hybrid" cross polytopes consist of the vertex-pairs (or rays) within the 18 outer circles in Fig. \ref{hex1}. Thus the geometrical origin of the 24 bases formed by the Peres rays should now be clear: three arise from the cross polytopes in Cell A, three from the cross polytopes in Cell B and the remaining 18 from the hybrid cross polytopes whose vertices originate equally from Cells A and B.\\

\begin{figure}[htp]
\begin{center}
\includegraphics[width=0.80\textwidth]{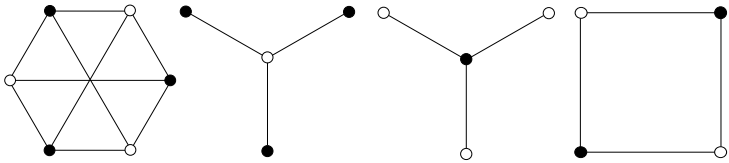}
\end{center}
\caption{Kochen-Specker diagrams of a Hexagon (left), a Trident (middle two figures) and a Square (right). Solid circles represent rays of one 24-cell and open circles those of its dual. The number of configurations of these four types within the 24 Peres rays is 16,48,48 and 144, respectively.}
\label{KS4}
\end{figure}

Any three rays of a 24-cell will be said to form a {\bf line} if the components of one can be expressed as a linear combination of those of the other two. The 12 rays of a 24-cell form 16 lines, with three rays lying on every line and four lines passing through every ray. This $12_{4}16_{3}$ system of rays and lines is an instance of a {\bf Reye's configuration} \cite{Hilbert}. The lines in a pair of dual 24-cells have the property that all the rays on a line of one are orthogonal to all the rays on a certain line of the other. Table \ref{Lines} shows the 16 lines of 24-cells A and B, with the orthogonal lines of the two shown next to each other. We will refer to a pair of orthogonal lines as a {\bf Hexagon} because the Kochen-Specker diagram \cite{KSdiag} of the six rays in them is a hexagon with all its edges and diameters drawn in (see leftmost diagram in Fig. \ref{KS4}). Hexagons are of interest because they allow a simple construction of all the 18-9 proofs to be given \cite{Aravind2000}: one simply needs to retain the bases not involving any of the rays in a Hexagon.

\begin{table}[ht]
\centering 
\begin{tabular}{|ccc | ccc |} 
\hline 
$\{1,   5,   9\} $ & $\leftrightarrow $ & $ \{16,  18,  23\}$ & $\{3 ,  5  , 11 \} $ & $ \leftrightarrow $ & $ \{ 14 ,18, 21\}$\\
$\{1,   6,   10  \} $ & $ \leftrightarrow $ & $ \{16,  17,  24\}$ & $\{3  , 6  , 12 \} $ & $ \leftrightarrow $ & $ \{ 14 , 17 , 22\}$\\
$\{1  , 7  , 11 \} $ & $ \leftrightarrow $ & $ \{ 15, 18, 24\}$&  $\{3 ,  7  , 9  \} $ & $ \leftrightarrow $ & $ \{ 13  ,18 , 22\}$\\
$\{1  , 8 ,  12  \} $ & $ \leftrightarrow $ & $ \{15 , 17 , 23\}$& $\{3  , 8  , 10 \} $ & $ \leftrightarrow $ & $ \{13, 17, 21\}$\\
$\{ 2  , 5  , 10 \} $ & $ \leftrightarrow $ & $ \{ 16 , 20 , 21\}$&  $\{ 4  , 5 ,  12 \} $ & $ \leftrightarrow $ & $ \{ 14 , 20,  23\}$\\
$\{2  , 6  , 9 \} $ & $ \leftrightarrow $ & $ \{  16, 19 ,22\}$& $\{4  , 6 ,  11 \} $ & $ \leftrightarrow $ & $ \{ 14 , 19 , 24\}$\\
$\{2  , 7  , 12 \} $ & $ \leftrightarrow $ & $ \{ 15 , 20 , 22\}$& $\{4  , 7 ,  10 \} $ & $ \leftrightarrow $ & $ \{13 ,20 ,24\}$\\
$\{2 ,  8 ,  11 \} $ & $ \leftrightarrow $ & $ \{ 15 , 19 , 21\}$& $\{4  , 8  , 9 \} $ & $ \leftrightarrow $ & $ \{  13 , 19 , 23\}$\\
\hline
\end{tabular}
\caption{Lines of 24-cell A matched with the corresponding (orthogonal) lines of 24-cell B. Each line is specified by the three rays on it}
\label{Lines} 
\end{table}

A line involves one ray from each of the bases within a 24-cell. Only 16 of the possible combinations of rays from the three bases lead to lines; the other 48 lead to entities we will term {\bf Triangles}. Each Triangle in a 24-cell has a unique ray in the dual 24-cell that is orthogonal to all the rays in it. We will refer to the combination of a Triangle and the ray orthogonal to it as a {\bf Trident} because the Kochen-Specker diagram of this set of four rays, shown in the middle two diagrams of Fig. \ref{KS4}, has the shape of this object. Tridents are of interest because they allow a simple construction to be given of one of the types of 20-11 proofs \cite{Aravind2000}: all one has to do is to retain the bases not involving any of the rays in a Trident. How does one obtain the remaining 20-11 proofs? Omitting a pair of rays belonging to different 24-cells from a Hexagon leaves four rays whose Kochen-Specker diagram is shown to the extreme right of Fig. \ref{KS4}, and which we term a {\bf Square}. Keeping all bases not involving any of the rays in a Square yields 12 bases from which one can be dropped to yield a 20-11 proof of the second type (the basis to be dropped is easily picked: it is the only one involving rays that all occur twice among the remaining 11 bases).\\

In summary, we have shown how to obtain all 18-9 and 20-11 proofs by keeping only bases not involving any of the rays that make up one of the four types of Kochen-Specker diagrams shown in Fig. \ref{KS4}. It is these prescriptions that were translated into the solutions of the puzzle presented earlier.\\

    We mention, for completeness, that the symmetry group of the Peres rays has 1152 elements in it. These may be accounted for as follows. Each of the 24-cells from which the Peres rays are derived has $3\cdot4!\cdot2^{4}=1152$ elements in its symmetry group {\cite{Coxeter}. Further, the symmetry groups of the two (dual) 24-cells coincide because they share the same symmetry planes. The Clifford displacement that exchanges the two 24-cells doubles the number of elements in the symmetry group to 2304. However, because the Peres rays exist in projective space, their symmetry group has only half the number of elements of the pair of 24-cells, or 1152.

\newpage

{\bf Ray-criticality versus basis-criticality}\\

Not all the parity proofs presented above are of the same type. Some are {\bf ray-critical} while others are {\bf basis-critical}. We now define these terms and point out the differences between them. A set of rays is said to be ray-uncolorable if one cannot assign the value 0 or 1 to each of them in such a way that (a) no two orthogonal rays are both assigned the value 1, and (b) not all members of a complete orthogonal set of rays (numbering four, in the present case) are assigned the value 0. If this value assignment proves possible, the set is said to be ray-colorable. A set of rays is said to be ray-critical if it is ray-uncolorable but the deletion of even a single ray from it makes the remaining rays ray-colorable.  A set of $R$ rays and $B$ bases, or a $R$-$B$ set, is said to be basis-uncolorable if the values 0 or 1 cannot be assigned to each of the rays in such a way that each basis has exactly one 1 and three 0's in it. A $R$-$B$ set is said to be {\bf basis-critical} if it is basis-uncolorable but the deletion of even a single basis from it makes the remaining rays and bases basis-colorable. Note the subtle difference in the requirements for ray-uncolorability and basis-uncolorability: ray-uncolorability takes into account {\bf all} orthogonalities between the rays considered whereas basis-uncolorability takes into account only the orthogonalities contained within the bases of interest (which generally do not include all the orthogonalities between the rays considered).  Ray-criticality of a set of rays does not guarantee that the bases made up solely of those rays are basis-uncolorable, let alone basis-critical. By contrast, basis-criticality of a set of rays and bases guarantees that the rays occurring in those bases are ray-uncolorable but not that they are ray-critical.\\

Let us refer to the 20-11 proofs obtained by deleting Tridents as 20-11A proofs and those obtained by deleting Squares as 20-11B proofs. It has been shown \cite{Aravind2000} that the only ray-critical sets of rays within the Peres set are the 18- and 20-ray sets obtained by deleting Hexagons or Tridents.  These 18- and 20-ray sets give rise to the 18-9 and 20-11A parity proofs that also happen to be basis-critical. The remaining proofs (20-11B, 22-13 and 24-15) are basis-critical but not ray-critical. These proofs can all be reduced to 18-9 or 20-11A proofs by shedding some bases and adding others in such a way that the rays finally involved are a subset of the original ones. We provide a couple of examples of how this can be done. Let us number the inner circles in Fig. \ref{hex1} from 1 to 6 and the outer circles from 7 to 24, beginning at the top and proceeding clockwise in both cases. Then the 22-13 proof of Fig. \ref{hex3}, which involves the bases 1,3,4,5,7,12,14,15,18,19,20,22 and 24, can be converted into a 20-11A proof by dropping the bases 1,5,7,14 and 19 and adding the bases 11,16 and 21. The 20-11A proof involves a subset of the rays of the 22-13 proof but not a subset of its bases, reflecting the fact that the latter proof is not ray-critical but is basis-critical. A different deletion and addition of bases can convert the same 22-13 proof into a 18-9 proof; one of the ways of doing this is to delete the bases 1,3,4,5,7,14 and 18 and add the bases 11,16 and 21. The 18-9 and 20-11A proofs are clearly the most interesting of our parity proofs because they are the only ones that cannot be reduced to smaller parity proofs involving a subset of their rays.\\

Our discussion has been confined entirely to parity proofs. However it should be pointed out that the Peres rays contain many sets of bases that provide proofs of the K-S theorem that are not of the parity type. These proofs may not be as appealing as parity proofs, as their verification takes more than simple counting, but they are just as conclusive. Many examples of such proofs are given in \cite{Pavicic2010}.\\

{\bf Relation to ongoing work}\\

The Peres rays were the first system in which parity proofs of the K-S theorem were discovered. It is interesting that there are four different types of parity proofs that come in complementary pairs (see Table \ref{Summary}), and that simple constructions for all of them can be given. Perhaps the only simpler proof of the K-S theorem is that due to Mermin \cite{Mermin1993}, which also uses the observables in Table \ref{Rays} but exploits them in a somewhat different way. Mermin arranges the observables in the form of a 3 x 3 square and shows that noncontextual values cannot be assigned to them in keeping with certain constraints. The Peres and Mermin proofs both spring from the same source, but are still different. If Mermin's proof is a bit like Alice in Wonderland, then Peres' is a bit like Through the Looking Glass. It is interesting that the Peres-Mermin proofs of the K-S theorem can be extended into proofs of Bell's nonlocality theorem if use is made of a maximally entangled state of a pair of qubits. This was demonstrated, in slightly different ways, in \cite{Cabello2001} and \cite{Aravind2002}. More recently it has been pointed out \cite{Aolita} how any proof of the K-S theorem can be extended into experimentally testable proofs of Bell's nonlocality theorem.\\

It is interesting to ask if there are other systems of rays that yield parity proofs of the K-S theorem. Kernaghan and Peres \cite{KP1995} found a set of 40 rays in 8 dimensions that yield parity proofs with the smallest known ratio of bases to rays in any dimension. More recently, two different sets of 60 rays have been found in four dimensions that yield parity proofs of the K-S theorem. The rays of one of set are derived from the vertices of a four-dimensional regular polytope known as the 600-cell \cite{Waegell2010}, while those of the other are derived from the bases associated with fifteen triads of commuting observables for a pair of qubits \cite{Waegell2011}. While the former rays are real, the latter are complex. Both these systems yield parity proofs of a wide range of sizes and types, with the total number being in the vicinity of a hundred million in each case. Cataloging all the different types of proofs, or even estimating their total number, is not an easy task. We therefore find it pleasing that it is possible to give a complete account of all the parity proofs in the Peres system. This is certainly a useful case to look at before exploring parity proofs in more complex systems.\\

Recently there has been renewed interest in the K-S theorem as a means for devising experimental tests of quantum contextuality. Cabello  \cite{Cabello2008} showed how to convert the 18-9 parity proof discussed above into an inequality that is obeyed by noncontextual hidden variables theories \cite{Gen}. Experimental tests of this and related inequalities have been carried out in four-state systems realized by ions \cite{Kirchmair}, neutrons \cite{Bartosik}, photons \cite{Amselem} and nuclear spins \cite{Moussa}, and violations of the inequalities have been observed in all the cases, showing that nature is contextual. In an interesting development, Klyachko et al \cite{Klyachko} have shown how the failure of the transitivity of implication for a closed loop of observables can be used to prove the K-S theorem. Two recent articles \cite{Liang,Bengtsson} expand on this theme and also explore the connection between the K-S theorem and a number of quantum paradoxes. In this connection it may be worth pointing out (see Appendix 2) that the Peres-Mermin square permits a simple state-independent demonstration to be given of the failure of the transitivity of implication for the observables occurring in it. A debate still continues about whether finite precision and compatibility loopholes negate experimental tests of the K-S theorem and demonstrations of contextuality \cite{Mayer}. It is possible that newer proofs of the K-S theorem \cite{Cunha} could suggest experiments that plug these loopholes and also find application in such protocols as quantum key distribution \cite{BPPeres}, random number generation \cite{Svozil} and parity oblivious transfer \cite{Spekkens}.\\

{\bf Acknowledgements.} We would like to thank Herman Servatius for useful discussions about the 24-cell and Paul Kassebaum for help with drafting some of the figures.\\

{\bf Appendix 1: a quick explanation of the K-S theorem}\\

The Kochen-Specker theorem rules out the existence of noncontextual hidden variables theories that have been proposed as deterministic alternatives to quantum mechanics. We give a simple explanation of what this statement means and how the K-S theorem accomplishes this goal. A more detailed exposition of the theorem can be found in \cite{PeresBook}.\\

One of the simplest types of measurement that can be carried out on a quantum system is a projective measurement. Such a measurement amounts to asking the system if it has a certain property or not, and the system replies with either a yes or a no (to which we attach the values 1 and 0, respectively). A set of projective measurements is said to be compatible if there exist special states of the system for which the measurements always yield the same values, no matter in what order they are carried out or how often they are repeated. A complete set of compatible projective measurements (CSCPM) is a maximal set of such measurements, i.e., one that cannot be enlarged by adding further measurements of the same type. For a pair of two-level systems (or qubits), which is one of the physical settings in which the Peres rays can be realized, the number of measurements in a CSCPM is four. An important property of a CSCPM is its exclusivity: whenever it is carried out on any state of a system, exactly one of the measurements returns the value 1 and all the others return the value 0. The properties of CSCPMs laid out so far are all well established experimental facts.\\

    How is one to interpret the results of a CSCPM? A realist, such as Einstein, who believes that the system possesses definite values of its properties even in the absence of measurement might maintain that the properties are merely revealed through the act of measurement. However orthodox quantum mechanics holds that the system does not generally possess definite values of all properties in a CSCPM (except in certain specially prepared states) and that the values only come into being during the act of measurement. How can one decide which of these viewpoints is right? That is what the K-S theorem helps us do, subject to a certain assumption.\\

   We now specialize the discussion to a system, such as a pair of qubits, for which the CSCPMs consist of four measurements each. For such a system it is always possible to choose 24 projective measurements, which we label 1-24, that can be grouped into the 24 CSCPMs shown in the circular boxes in Fig.\ref{hex1}; in other words, the projective measurements and CSCPMs can be made to correspond to the Peres rays and their bases, respectively. A realist, who believes that the values observed for a CSCPM exist even before measurement, and who also subscribes to the assumption of noncontextuality (i.e., the notion that the value observed for a particular projective measurement is independent of the CSCPM it is carried out as a part of) would be faced with the task of assigning a 0 or a 1 to each of the Peres rays in such a way that each basis contains exactly one 1 and three 0's in it. However the various parity proofs presented in this paper demonstrate that this task is impossible, and so the realist's position is refuted. The key assumption made by the realist that undermines his position is the assumption of noncontextuality. If this assumption is abandoned in favor of the notion of locality (i.e., that physical influences cannot propagate faster than light), the realist's position can still be refuted by an argument due to Bell \cite{Bell1964} or any of its later variations.\\

{\bf Appendix 2: failure of transitivity of implication for the Peres-Mermin square}\\

The upper part of Fig. \ref{Trans} shows the 9 observables of the Peres-Mermin square. Each observable can assume only the values +1 or -1 and the product of the values in any row or column must have the value indicated to its right or bottom. It is convenient to replace the values +1 and -1 by 0 and 1, respectively, and the constraints are then that the sum of the values in any row or column, modulo 2, must equal 0, except for the sum of the values in the third row, which must equal 1. The argument now proceeds as shown in the four diagrams below the top square (with the diagrams being traversed clockwise, starting from the one at the top left): if arbitrary values $a,b,c$ and $d$ are assumed for four of the observables, the transitivity of implication (depicted by the arrows in the figure) shows how the values of the remaining observables are fixed by the row and column constraints; finally one returns to the top left corner of the square with the value $a+1$, which contradicts the value $a$ that one began out with.

\begin{figure}[htp]
\begin{center}
\includegraphics[width=0.80\textwidth]{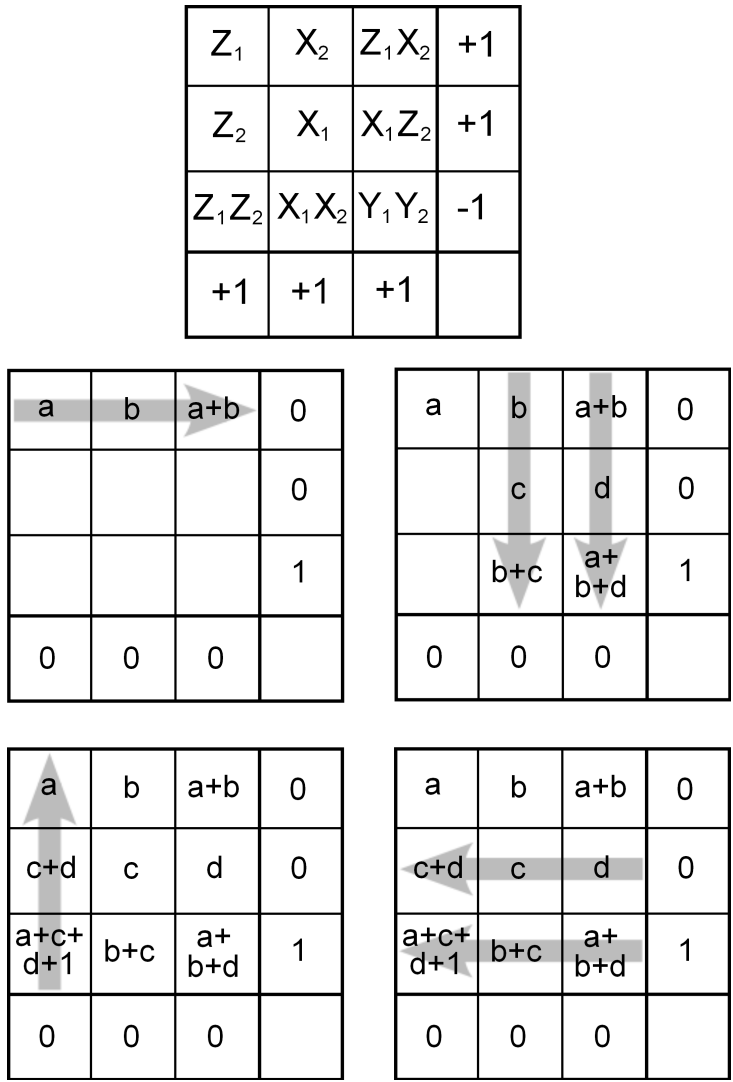}
\end{center}
\caption{Failure of transitivity of implication for the observables in the Peres-Mermin square.}
\label{Trans}
\end{figure}


\clearpage

\begin{thebibliography}{10}
\providecommand{\url}[1]{{#1}}
\providecommand{\urlprefix}{URL }
\expandafter\ifx\csname urlstyle\endcsname\relax
  \providecommand{\doi}[1]{DOI \discretionary{}{}{}#1}\else
  \providecommand{\doi}{DOI \discretionary{}{}{}\begingroup
  \urlstyle{rm}\Url}\fi

\bibitem{Peres1991}
A.~Peres: {\it J. Phys. A} \textbf{{\bf 24}}, L175 (1991).

\bibitem{KS1967}
S.~Kochen, E.P.~Specker: {\it J. Math. Mech.} \textbf{{\bf 17}}, 59 (1967). See also J.S.~Bell: {\it Rev. Mod. Phys.} \textbf{{\bf 38}}, 447 (1966).
\newblock Reprinted in J.S.~Bell: {\em{Speakable and Unspeakable in Quantum
  Mechanics}}, (Cambridge University Press, Cambridge, 1987).

\bibitem{Kernaghan1994}
M.~Kernaghan: {\it J. Phys. A} \textbf{{\bf 27}}, L829 (1994)

\bibitem{Cabello1996}
A.~Cabello, J.M.~Estebaranz, G.~{Garc{\'\i}a-Alcaine}: {\it Phys. Lett. A}
  \textbf{{\bf 212}}, 183 (1996)

\bibitem{Pavicic2010}
M.~Pavi{\v c}i{\'c}, N.D. Megill, J.P. Merlet: {\it Phys. Lett. A} \textbf{{\bf
  374}}, 2122 (2010); M.~Pavi{\v c}i{\'c}, J.P. Merlet, B.D. Mc{K}ay, N.D. Megill: {\it J. Phys. A}
  \textbf{{\bf 38}}, 1577 (2005); M.~Pavi{\v c}i{\'c}, J.P. Merlet, N.D. Megill: {\it The {F}rench {N}ational
  {I}nstitute for {R}esearch in {C}omputer {S}cience and {C}ontrol {R}esearch
  {R}eports} \textbf{{\bf {RR-5388}}} (2004).

\bibitem{Coxeter}
H.~Coxeter: \emph{Regular Polytopes} (Dover, New York, 1973). The transformation (\ref{Clifford}), applied four times, gives inversion through the origin and, applied eight times, the identity.

\bibitem{Hilbert}
D.~Hilbert and S. Cohn-Vossen: \emph{Geometry and the Imagination} (Chelsea, New York, 1983).

\bibitem{KSdiag}
A Kochen-Specker diagram shows all the orthogonalities within a set of rays by representing the rays as points and drawing lines between all pairs of points that represent orthogonal rays.

\bibitem{Aravind2000}
P.K.~Aravind: {\it Found. Phys. Lett.} \textbf{{\bf 13}}, 499 (2000).

\bibitem{Mermin1993}
N.D.~Mermin: {\it Phys. Rev. Lett.} \textbf{{\bf 65}}, 3373 (1990); {\it Rev. Mod. Phys.} \textbf{{\bf 65}}, 803 (1993).

\bibitem{Cabello2001}
A.~Cabello: {\it Phys. Rev. Lett.}  \textbf{{\bf 86}}, 1911 (2001); {\it Phys. Rev. Lett.} \textbf{{\bf 87}}, 010403 (2001).

\bibitem{Aravind2002}
P.K.~Aravind: {\it Found. Phys. Lett.} \textbf{{\bf 15}}, 399 (2002).

\bibitem{Aolita}
L.~Aolita, R. Gallego, A. Acin, A. Chiuri, G. Vallone, P. Mataloni, A. Cabello: ArXiv:1105.3598 (2011).

\bibitem{KP1995}
M.~Kernaghan, A. Peres: {\it Phys. Lett.} \textbf{{\bf A198}}, 1 (1995).

\bibitem{Waegell2010}
M.~Waegell, P.K. Aravind: {\it J. Phys. A} \textbf{{\bf 43}}, 105304 (2010); M. Waegell, P.K. Aravind, N.D. Megill, M.~Pavi{\v c}i{\'c}: {\it Found. Phys.} \textbf{{\bf 41}}, 883 (2011).

\bibitem{Waegell2011}
M.~Waegell and P.K. Aravind: unpublished.

\bibitem{Gen}
 A generalization of this argument to arbitrary parity proofs of the K-S theorem in any even dimension was given in the second paper in \cite{Waegell2010}.

\bibitem{Cabello2008}
A.~Cabello: {\it Phys. Rev. Lett.} \textbf{{\bf 101}}, 210401 (2008).
\newblock See also: P. Badzi{\c a}g, I. Bengtsson, A. Cabello, I. Pitowsky: {\it
  Phys. Rev. Lett.} {\bf 103} (2009) 050401

\bibitem{Kirchmair}
G.~Kirchmair, F.~Z{\"a}hringer, R.~Gerritsma, M.~Kleinmann, O.~G{\"u}hne,
  A.~Cabello, R.~Blatt, C.F.~Roos: {\it Nature} \textbf{{\bf 460}}, 494 (2009)

\bibitem{Bartosik}
H.~Bartosik, J.~Klep, C.~Schmitzer, S.~Sponar, A.~Cabello, H.~Rauch,
  Y.~Hasegawa: {\it Phys. Rev. Lett.} \textbf{{\bf 103}}, 040403 (2009)

\bibitem{Amselem}
E.~Amselem, M.~R{\aa}dmark, M.~Bourennane, A.~Cabello: {\it Phys. Rev. Lett.}
  \textbf{{\bf 103}}, 160405 (2009)

\bibitem{Moussa}
O.~Moussa, C.A. Ryan, D.G. Cory, R.~Laflamme: {\it Phys. Rev. Lett.}
  \textbf{{\bf 104}}, 160501 (2010)

\bibitem{Klyachko}
A.A.~Klyachko, M.A. Can, S. Binicio\u{g}lu and A.S. Shumovsky: {\it Phys. Rev. Lett.}
  \textbf{{\bf 101}}, 020403 (2008)

\bibitem{Liang}
Y.C.~Liang, R.W. Spekkens, H.M. Wiseman: {\it {A}r{X}iv:1010.1273v1} (2010)

\bibitem{Bengtsson}
P.~Badzi{\c a}g, I. Bengtsson, A. Cabello, H.H. Granstr\"om and J.-\AA. Larsson: {\it Found Phys}
  \textbf{{\bf 41}}, 414 (2011)

\bibitem{BPPeres}
H.~Bechmann{-P}asquinucci, A.~Peres: {\it Phys. Rev. Lett.} \textbf{{\bf 85}},
  3313 (2000); K. Svozil: ArXiv:0903.0231 (2009)

\bibitem{Svozil}
K.~Svozil: {\it Phys. Rev. A} \textbf{{\bf 79}}, 054306 (2009)

\bibitem{Spekkens}
R.W.~Spekkens, D.H.~Buzacott, A.J.~Keehn, B.~Toner, G.~Pryde: {\it Phys. Rev.
  Lett.} \textbf{{\bf 102}}, 010401 (2010)

\bibitem{PeresBook}
A.~Peres: {\em{Quantum Theory: Concepts and Methods}}, (Kluwer Academic, Dordrecht, 1993).

\bibitem{Bell1964}
J.S.~Bell: {\it Physics} \textbf{{\bf 1}}, 195 (1964), reprinted in the book mentioned in Ref.2.

\bibitem{Mayer}
D.A.~Mayer: {\it Phys. Rev. Lett.} \textbf{{\bf 83}}, 3751 (1999); A.~Kent: {\it Phys. Rev. Lett.} \textbf{{\bf 83}}, 3755 (1999);
J.~Barrett and A.~Kent, {\it Stud. Hist. Philos. Mod. Phys.} \textbf{{\bf 35}}, 151 (2004); N.D.~Mermin, quant-ph/9912081. 

\bibitem{Cunha}
A.~Cabello and M.~Terra Cunha: Phys. Rev. Lett. \textbf{{\bf 106}}, 190401 (2011).

\end{thebibliography}

\end{document}